\documentclass{article}  
\usepackage{fix-cm}
\usepackage{authblk}
\usepackage{graphicx}

\author[1,*]{C. Sailer}
\author[2]{B. Lubsandorzhiev}
\author[1]{C. Strandhagen}
\author[1]{J. Jochum}

\affil[1]{Kepler Center for Astro and Particle Physics, Auf der Morgenstelle 14, 72076 T\"ubingen, Germany}
\affil[2]{Institute for Nuclear Research, Russian Academy of Sciences, pr. Shestidesyatiletiya Oktyabrya 7a, Moscow, Russia}

\begin{document}

\title{Low temperature light yield measurements in NaI and NaI(Tl)
}
\maketitle

\let\oldthefootnote\thefootnote
\renewcommand{\thefootnote}{\fnsymbol{footnote}}
\footnotetext[1]{Corresponding author: sailer@pit.physik.uni-tuebingen.de}
\let\thefootnote\oldthefootnote

\begin{abstract}
The scintillation light output of a pure and a Thallium doped Sodium Iodide (NaI) crystal under irradiation with 5.486\,MeV $\alpha$ particles has been measured over a temperature range from 1.7\,K to 300\,K. Estimates of the decay time constant at three selected temperatures are given. For pure NaI an increase in light yield towards low temperatures could be confirmed and measured at higher precision. For NaI(Tl) below 60\,K an increase in light output has been found.\\

\end{abstract}

\section{Introduction}
\label{intro}
In recent years scintillating crystals at low temperatures used as bolometers have gained interest due to their application in rare event searches, such as the CRESST Dark Matter Search~\cite{RefCresst} and others, where particle identification is required.  
Employing different target materials is of special interest in direct Dark Matter Experiments. It allows one to scan the anticipated atomic number A$^2$ dependence of the WIMP-nucleon cross section, that should be present if the scattering is a coherent process. Different materials are investigated at the moment. Here we report on NaI.\\
As van Sciver and Bogart~\cite{RefSciver} and Collinson and West~\cite{RefWest} 
reported, the light yield of pure NaI increases towards lower temperatures. As low temperature bolometers are operated well below liquid helium temperature (e.g. at mK  range) the effort of this work was to extend available data down to 1.7\,K and verify the results reported in~\cite{RefSciver} and~\cite{RefWest}. As a reference a Thallium doped NaI sample of the same geometry was also investigated. 
\section{Experimental setup}
\label{sec:exp}
The two crystal samples used in this work were produced by Saint-Gobain Crystals. Both samples are cubic shaped with an edge length of 1\,cm where one crystal is doped with Thallium and the other one is made of pure NaI. To ensure that the pure NaI sample was not accidentally doped with Thallium remnants, it was grown in a dedicated mold.\\
As NaI is highly hygroscopic the crystals were placed in a copper housing with quartz glass windows manufactured by Aachener Quarz-Glas Technologie Heinrich. The copper holder was assembled in a nitrogen purged glove box and sealed with indium wire. To prevent the remaining nitrogen inside of the capsule from condensing on the optical windows, a custom made cryogenic pump utilizing charcoal was added. An $^{241}$Am source with an activity of 840\,kBq was placed inside the housing to irradiate the crystal. 

\begin{figure}[hptb]
  \includegraphics[width=0.90\textwidth]{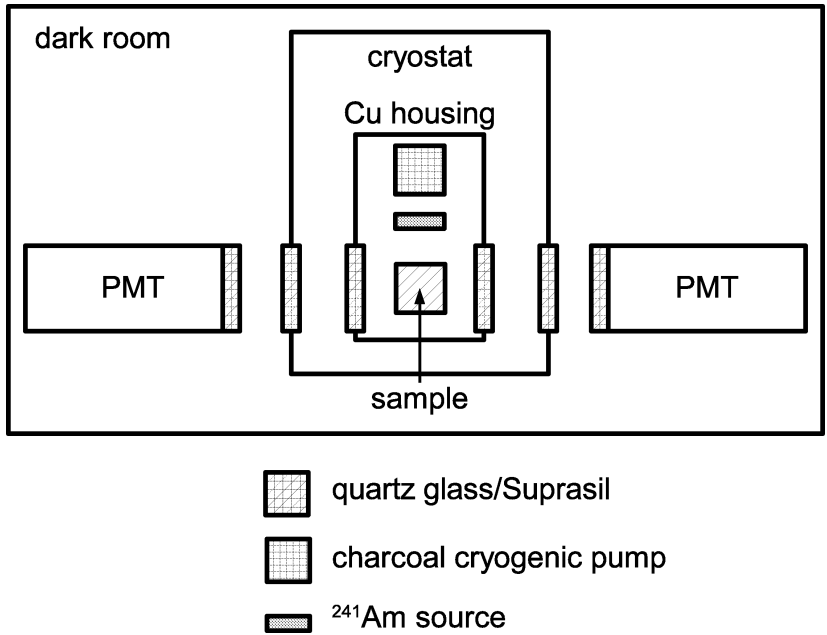}
\caption{Sketch of the experimental setup}
\label{fig:setup} 
\end{figure}

A Janis Research cryostat, model SVT-400, was used to cool down the crystals. The cryostat is equipped with indium-sealed Suprasil~II windows transmitting more than 90\,\% of the light for wavelengths greater than 190\,nm. This setup allows using photomultiplier tubes (PMTs) at room temperature, while the sample temperature can be sweeped. The sample is cooled by direct contact with vaporized liquid helium providing temperatures as low as 1.7\,K. 
In order to measure and stabilize the temperature, the cryostat is equipped with two DT-670B-SD diodes by Lakeshore, one measuring the bath temperature at the vaporizer, the other one mounted on the sample holder. A 25\,$\mathrm{\Omega}$ heater on the sample holder is connected to a Lakeshore~331S Temperature Controller, allowing to stabilize temperatures to a precision of up to 0.01\,K or better, depending on the temperature range.\\
As the maximum of the spectral emission of pure NaI shifts to 300\,nm at lower temperatures~\cite{RefWest}, type 9235\,QB quartz glass tubes produced by ET Enterprises were used. These PMTs provide a high quantum efficiency $\geq$ 25\,\% for $\lambda$ between 180\,nm and 440\,nm.\\
The PMT signals were fed into Ortec TF~474 amplifiers where they were amplified and integrated for 20\,ns. The signal was duplicated in a LeCroy 428F Fan~In - Fan~Out module (FiFo) and connected to a CAEN N840 leading edge discriminator with low threshold of a few photoelectrons that was used to get a trigger signal. The respective NIM logical pulses were then fed into a LeCroy 622 Quad Coincidence module to realize a logical AND trigger for coincident signals in the two PMTs. The amplified signal from the FiFo and the trigger pulse were connected to a SIS 3301 FADC that recorded the traces of the scintillation events with a sampling rate of 100\,MS/s. For every triggered event the software also recorded both temperature readings. Figure~\ref{fig:setup} shows a sketch of the setup.  

\section{Data analysis}
\label{sec:data}

\begin{figure}[hptb]
  \includegraphics[width=0.90\textwidth]{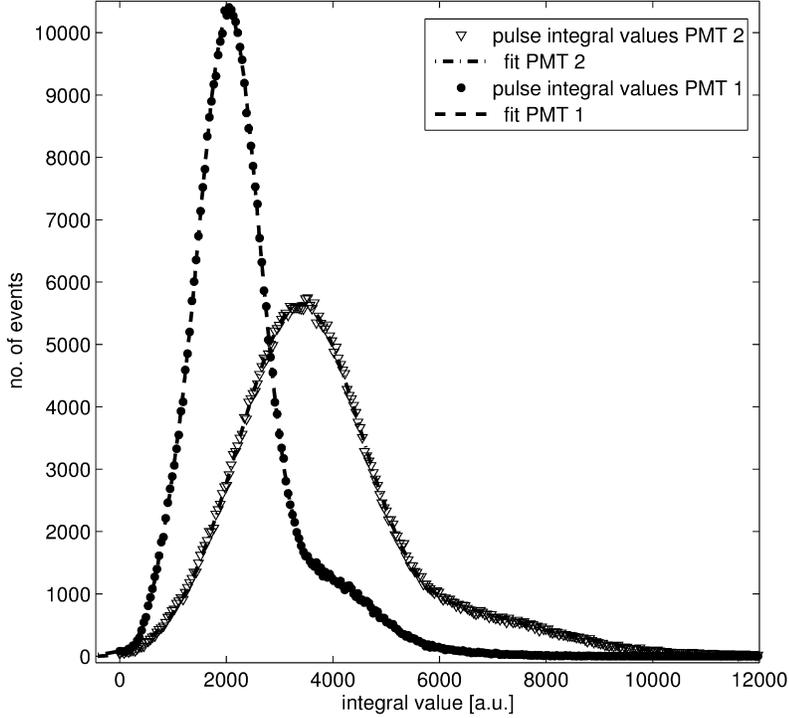}
\caption{Exemplary histogram of integral values for scintillation events seen by PMT\,1 and PMT\,2 at 178\,K. The fit accounts for baseline noise and the two $\alpha$~distributions (see text for details)}
\label{fig:spectrum}
\end{figure}

At each temperature 512,000 scintillation events were recorded from the NaI(Tl) and 1,024,000 from the pure NaI. A cut was applied to remove traces with nonzero pretrigger mean values. This removes roughly 7\,\% of the total events. The light output is proportional to the integrated PMT signals. Figure~\ref{fig:spectrum} shows an exemplary spectrum in pure NaI observed by PMT\,1 and PMT\,2 at a temperature of 178\,K. At higher values the peak exhibits a shoulder which can be attributed to double $\alpha$~hits during one record length. These events were not removed by a cut. Rather a fit with three Gaussians was applied to account for baseline noise and both $\alpha$~distributions. The positions of the maximum of the single $\alpha$~distribution were determined for both channels separately and the resulting values were added. Statistical errors are omitted, as they are negligible. Systematical errors have been derived from variation of the fit, using the 95\,\% confidence interval for the peak position of the single $\alpha$~peak. The two PMTs were not operated at equal gain due to limitations in the electronics of PMT\,2. The gain of PMT\,1 was set to a higher value as compared to PMT\,2 in order to increase the trigger rate. As all given values are relative quantities, this does not affect the results.\\  

The decay time constants of the signals from the two samples is different, which could already be noticed on the oscilloscope. The pure NaI was recorded with 2.56\,$\mu$s record length, the NaI(Tl) data with 5.12\,$\mu$s. To illustrate the temperature dependence, a mean decay time constant was derived in three representative temperature regimes for each crystal. This was done by selecting a sample of scintillation events with a cut on the ratio between pulse height and integral to prevents double events from faking long decay time constants. For each sample a summed pulse was then fitted with an exponential decay time. Note that due to the time resolution of the FADC used, fast components in the order of several nanoseconds can not be resolved. Also the 20\,ns shaping time in the preamplifier will slightly affect the values for $\tau$. 

\section{Results}
\label{sec:results}

\begin{figure}[hptb]
  \includegraphics[width=0.90\textwidth]{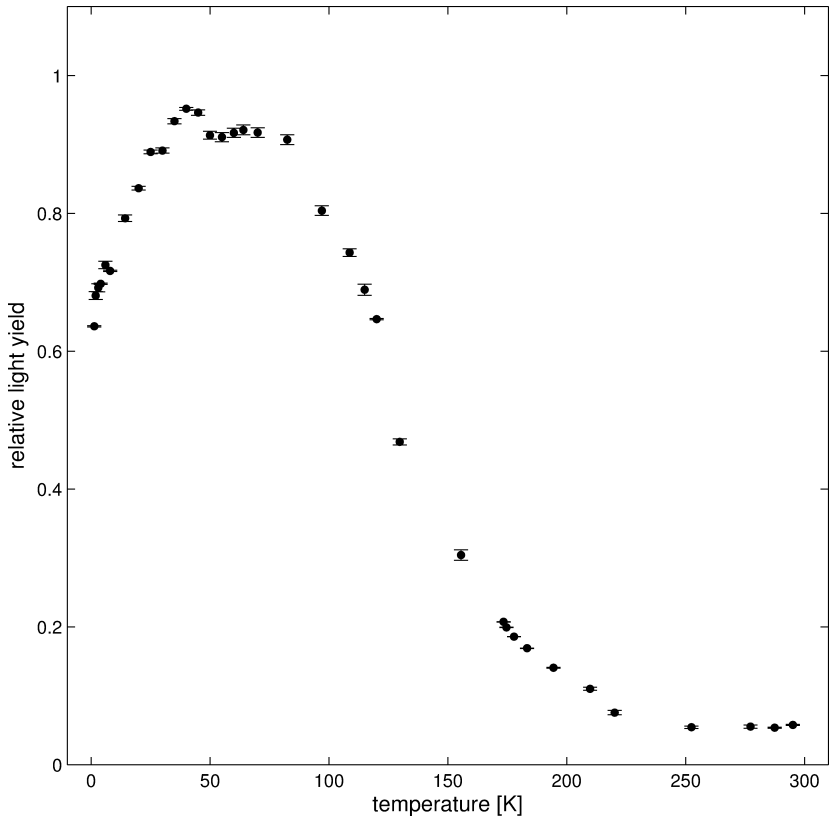}
\caption{Light yield of pure NaI as function of temperature relative to the value of NaI(Tl) at room temperature. Errors are dominated by systematics, see text. }
\label{fig:lyPure}
\end{figure}

\begin{figure}[hptb]
  \includegraphics[width=0.90\textwidth]{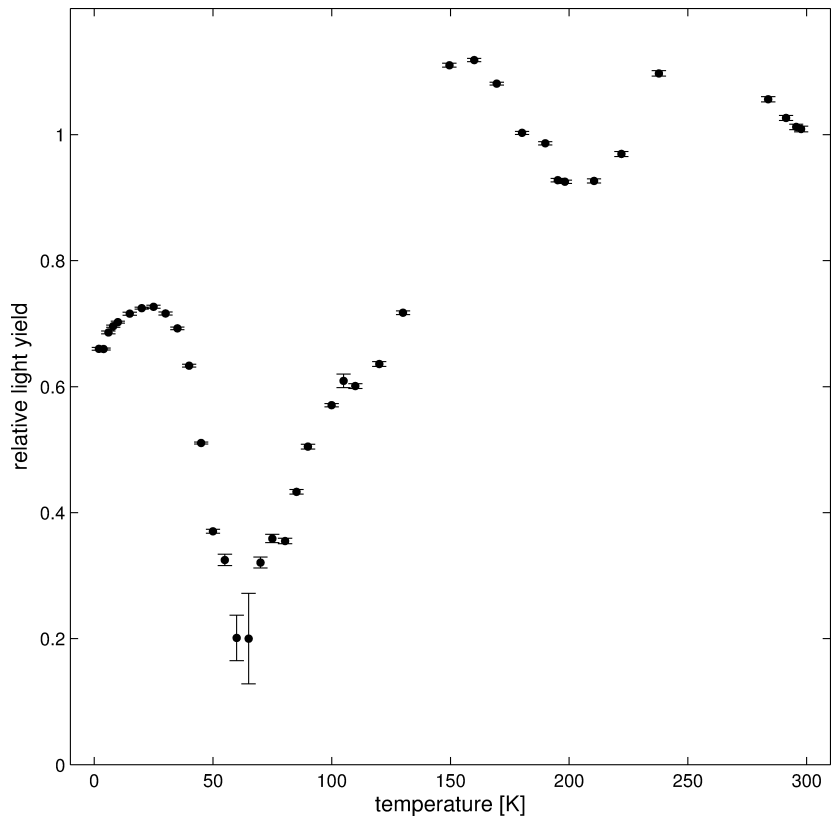}
\caption{Light yield of NaI(Tl) as function of temperature relative to the value at room temperature. Errors are dominated by systematics, see text for details. }
\label{fig:lyTl}
\end{figure}

Figure~\ref{fig:lyPure} shows the results for the light yield of pure NaI, Figure~\ref{fig:lyTl} the data for NaI(Tl) respectively. The light yield axes are scaled such that the room temperature value of NaI(Tl) is set to one.\\

For pure NaI the general light yield curve published by van Sciver and Bogart~\cite{RefSciver} is confirmed by our measurement. In this work, data has been slightly extended towards lower temperatures to examine the decrease in light output after the maximum at about 50\,K. The temperature resolution has also been increased. It is striking that the maximum value of pure NaI at around 50\,K is almost comparable to the light yield of Thallium doped NaI at room temperature. Van Sciver and Bogart reported approximately double the room temperature value of NaI(Tl), peaking at nitrogen temperature. As there has been quite some progress in producing higher light output NaI(Tl) crystals since the 1960s the used reference crystal in this work might exhibit a higher light yield, thus giving lower relative values. Also note that in \cite{RefSciver} the pulse height was used to deduce the light yield as compared to the integral in this work to take into account the varying decay time. Thus, the maximum of the light emission is found at higher temperatures as compared to Figure \ref{fig:lyPure}. At 1.7\,K the light yield of pure NaI amounts to approximately 65\,\% of the room temperature value of NaI(Tl). \\
For NaI(Tl) a strong suppression in light output is observed down to 60\,K. This has been known for quite some time \cite{RefBicron}. However we observe a substantial increase in light yield when lowering the temperature further. At 2.0\,K we obtain a value of roughly 65\,\% relative to the room temperature value of NaI(Tl) which is comparable to the results obtained for pure NaI.\\ 
This raises the question if the increase of light output below 60\,K for NaI(Tl) could be related to the intrinsic properties of pure NaI or it is due to another scintillation mechanism. To check this, the decay times at three temperatures have been derived as described in Section~\ref{sec:data}. Results can be found in Table~\ref{tab:taus}.

\begin{table}
\centering
\caption{Decay times derived from exponential fit}
\label{tab:taus}
\begin{tabular}{llll}
\hline\noalign{\smallskip}
NaI pure & & & \\
\noalign{\smallskip}\hline\noalign{\smallskip}
T & 290\,K & 156\,K & 6\,K \\
\noalign{\smallskip}
$\tau_1$ & 98\,ns & 112\,ns & 149\,ns \\
$\tau_2$ & - & 644\,ns & - \\
\noalign{\smallskip}\hline\noalign{\smallskip}
NaI(Tl) & & & \\
\noalign{\smallskip}\hline\noalign{\smallskip}
T & 300\,K & 150\,K & 6\,K \\
\noalign{\smallskip}
$\tau$ & 219\,ns & 752\,ns & 115\,ns \\
\noalign{\smallskip}\hline
\end{tabular}
\end{table} 

For pure NaI the decay time slightly increases towards lower temperatures. Van Sciver and Bogart reported the same tendency but give much smaller values, especially at room temprerature~\cite{RefSciver}. Yet Pooley and Runciman~\cite{RefPooley} state that those decay time values might not be reliable due to trace impurity concentrations. At 156\,K the pulse can only be fitted when introducing a second decay time. For consistency we therefore used a double exponential model for all values of pure NaI but found that only at 156\,K the second exponential gives nonzero values.\\
For the decay time of NaI(Tl) at  300\,K we find 219\,ns which is in very good agreement with the literature value~\cite{RefKnoll} of 230\,ns at 293\,K which is why we consider the values in this work reliable. Decreasing the temperature of the NaI(Tl) sample leads to longer decay times peaking around 150\,K. Quite unexpectedly at 6\,K the pulses are faster again, even faster than at room temperature. 
The measured 115\,ns are close to the value of 110\,ns for intrinsic luminescence in pure NaI reported in Fontana et al.~\cite{RefFontana} for band to band excitation. So it seems possible, that at low temperatures the light yield of NaI(Tl) is dominated by the intrinsic properties of the NaI lattice. This seems to support the conclusion of Fontana et al. that "at high temperatures the scintillation time constant seems then to be mainly determined by the energy transfer mechanism, and not by the intrinsic behavior of the luminescence center"~\cite{RefFontana}.\\      

\section{Conclusion}
\label{sec:conclusion}

In this work we have confirmed that the light yield of pure NaI increases at low temperatures as reported by van Sciver and Bogart~\cite{RefSciver}. The data was slightly extended towards lower temperatures and resolved with higher precision than before. At 1.7\,K the light yield of pure NaI still amounts to about 65\,\% of the light output of NaI(Tl) at room temperature. \\
For NaI(Tl) an increase in light output has been found below 60\,K which, to our knowldege has not been reported yet and might be possibly attributed to intrinsic luminescence of NaI. At 2\,K for NaI(Tl) also a value of about 65\,\% of the room temperature light yield is found.

\section*{Acknowledgements}
The authors would like to thank Claudia Osswald for setting up the cryostat gas handling system. This work was partly funded by SFB-TR~27.

\end{document}